\documentclass[epj]{svjour}
\usepackage{graphics}
\newcommand{\bm}[1]{\mbox{\boldmath $#1$}}
\begin{document}
\title{Future of low-x forward physics at RHIC}
\author{
L.C. Bland\inst{1} \and 
F. Bieser\inst{2} \and
R.L. Brown\inst{1} \and
H.J. Crawford\inst{2} \and
A.A. Derevshchikov\inst{4} \and
J.L. Drachenberg\inst{5} \and
J. Engelage\inst{2} \and
L. Eun\inst{3} \and
C.A. Gagliardi\inst{5} \and
S. Heppelmann\inst{3} \and
E.G. Judd\inst{2} \and
V.I. Kravtsov\inst{4} \and
Yu.A. Matulenko\inst{4} \and
A.P. Meschanin\inst{4} \and
D.A. Morozov\inst{4} \and
L.V. Nogach\inst{4} \and
S.B. Nurushev\inst{4} \and
A. Ogawa\inst{1} \and
C. Perkins\inst{2} \and
G. Rakness\inst{1,3} \and
K.E. Shestermanov\inst{4} \and
A.N. Vasiliev\inst{4}}
%
%
\institute{
Brookhaven National Laboratory \and
University of Berkeley/Space Sciences Institute \and
Pennsylvania State University \and
IHEP, Protvino \and
Texas A\&M University}
\date{Received: date / Revised version: date}
%
\abstract{
Comparisons of particle production from high-energy ion collisions
with next-to-leading order perturbative QCD calculations show good
agreement down to moderate transverse momentum values.  Distributions
of azimuthal angle differences between coincident hadrons in these
collisions support a partonic origin to the particle production, again
down to moderate transverse momentum values.  The rapidity dependence
of inclusive and coincident particle production can therefore be used
to probe parton distribution functions down to small momentum
fractions where theory anticipates that parton saturation could be
present.  This paper describes how such experiments could be
completed.}
\PACS{
      {12.38 Qk}{}
      \and
      {13.88.+e}{}
      \and
      {24.85.+p}{}
     } 
%
\maketitle
\section{Introduction}
\label{intro}

Comprehensive measurements of p+p, d+Au and Au+Au interactions at
$\sqrt{s_{NN}}$=200 GeV by the RHIC experiments strongly suggest that the
central collisions of two gold nuclei lead to a new form of matter
that appears opaque to high transverse momentum ($p_T$) hadrons
\cite{RHIC}.  This dense matter evolves from an initial state produced
by the collisions of the low-$x$ gluon fields of each nucleus
\cite{GM}.  Understanding this initial state is the first topic
mentioned in a recent report on scientific opportunities in heavy-ion
physics.  ``Upgraded forward instrumentation'' was identified as
needed to elucidate the properties of the initial state \cite{barnes}.
Nucleon gluon density distributions are determined by global fits to
data \cite{CTEQ,MRST}, but the low-$x$ nuclear gluon distribution is
not yet known \cite{HKM,CERN_pA}.  The nuclear gluon field
distribution might be naively expected to result from a convolution of
the gluon density distributions of all the individual nucleons.
However, there is indirect experimental evidence from RHIC
\cite{BRAHMS} that the low-$x$ gluon distribution in a large nucleus
like gold is reduced, or shadowed, from the nominal superposition of
the distributions of the included protons and neutrons, a phenomenon
described as saturation.  In this document we describe a possible
measurement of the gluon distribution in a large nucleus.

\begin{figure}
\begin{center}
\resizebox{0.83\hsize}{!}{\includegraphics{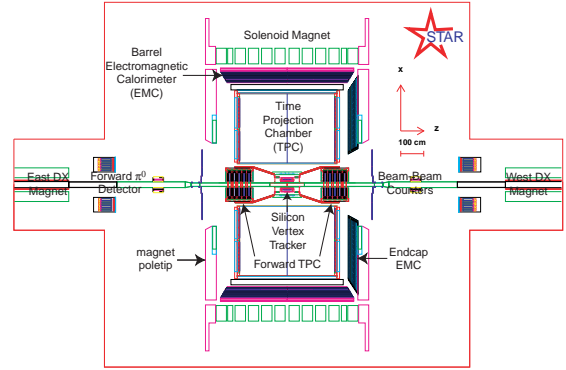}}
\end{center}
\caption{Layout of the Solenoidal Tracker at RHIC \cite{STAR}}
\label{STAR}
\end{figure}

\begin{figure}
\begin{center}
\resizebox{0.35\hsize}{!}{\includegraphics{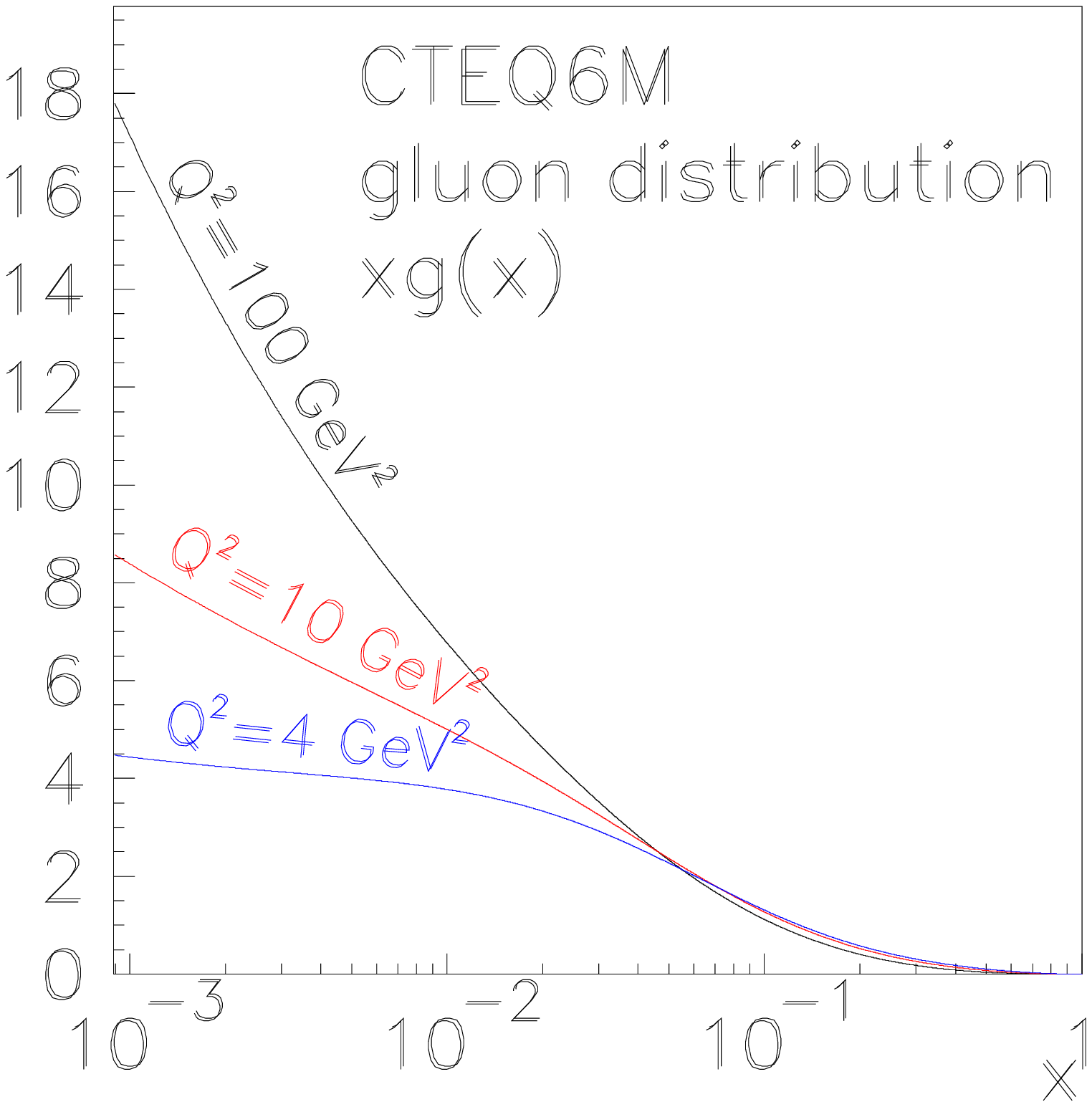}}
\resizebox{0.63\hsize}{!}{\includegraphics{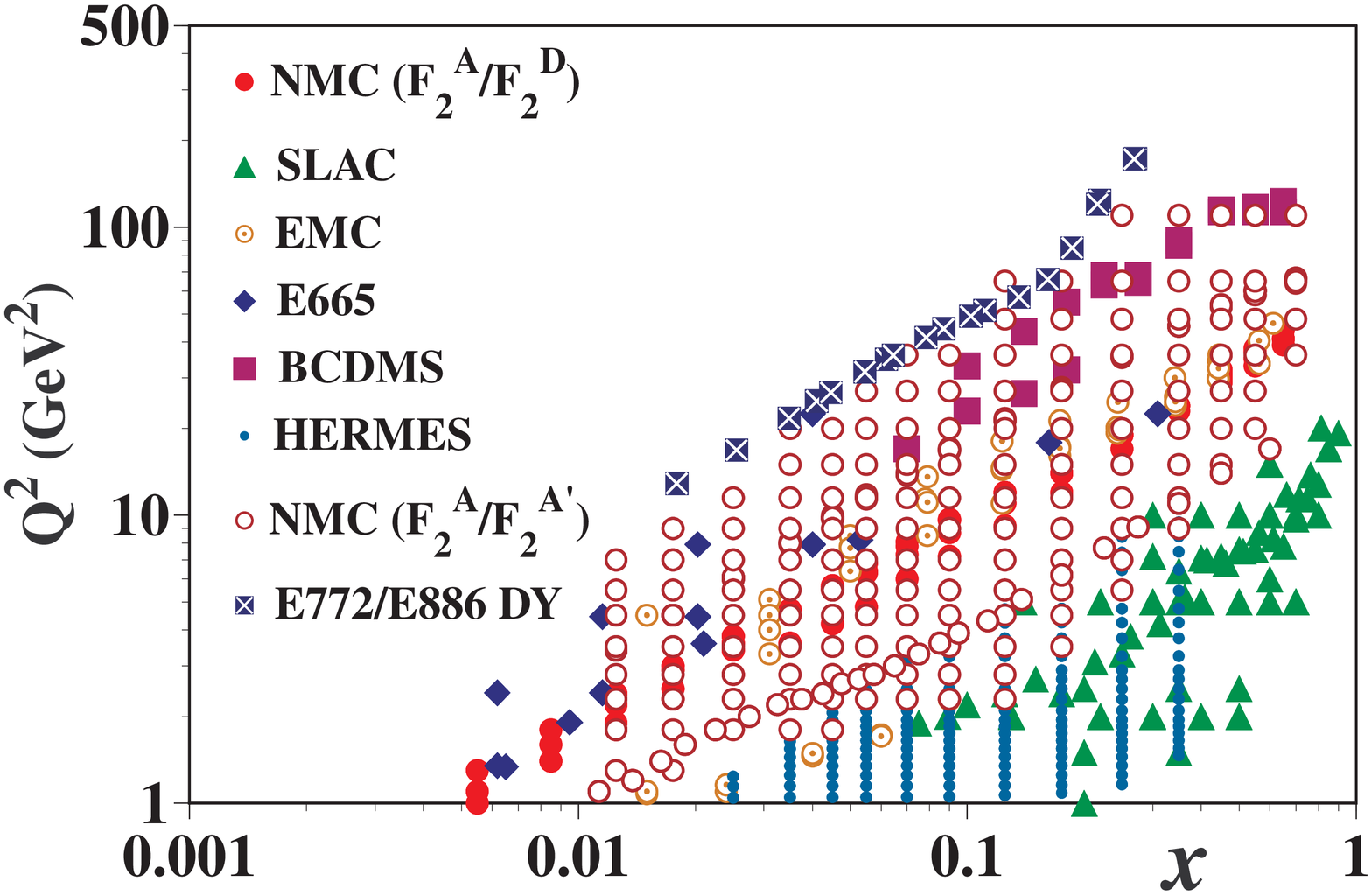}}
\end{center}
\caption{(Left) The gluon distribution in the proton
  \cite{CTEQ}.  Note the rapid rise in $xg(x)$ for $x<0.01$, a
  discovery made at HERA based on studies of deep inelastic scattering
  (DIS), using electron(positron)+proton collisions at $\sqrt{s}$=300
  GeV \cite{H1_gluon,ZEUS_gluon}.  (Right) Values of DIS kinematic
  variables $x$ and $Q^2$ where nuclear data from fixed-target
  DIS experiments at much lower $\sqrt{s}$ constrain the nuclear
  gluon density for $x>0.02$ \cite{HKM,CERN_pA}.}
\label{fig:1}       
\end{figure}

In the early runs at RHIC, we have demonstrated that forward
electromagnetic calorimeters (Fig.~\ref{STAR}), 
the Forward $\pi^0$ Detector (FPD), can be used to measure
$\pi^0$ produced at large pseudorapidity ($\eta=-\ln\tan\theta/2$,
where $\theta$ is the polar angle of the produced particle) in p+p and
d+Au collisions at $\sqrt{s_{NN}}$=200 GeV at STAR \cite{STAR_FPD}.
We have recently proposed to assemble a Forward Meson Spectrometer
(FMS) that will be operated during future RHIC running periods by the
STAR collaboration to enable measurement of the gluon distribution,
$xg(x)$, in nuclei in the range $0.001<x<0.1$.  The function $g(x)$
gives the differential probability to find gluons with a fraction $x$
of the longitudinal momentum of the parent nucleon
(Fig.~\ref{fig:1}).  The FMS will cover the range $2.5<\eta<4.0$ and
give STAR nearly hermetic electromagnetic coverage in the range
$-1<\eta<4$.  The FMS will allow correlation measurements between
forward mesons and photons with signals from the full STAR detector,
including the barrel and endcap electomagnetic calorimeters
(BEMC,EEMC) and the forward and midrapidity time-projection chambers
(TPC).  Exploiting the capabilities of RHIC and the existing STAR
detector, and assuming simple 2-body kinematics, the FMS will allow
measurement of the gluon density in protons and in nuclei down to
$x\sim0.001$.

With the addition of the FMS, which has $>25$ times larger areal
coverage than the FPD we view as its prototype, STAR will be able to
achieve at least three important and new physics objectives:
\begin{itemize}
\item
A measurement of the gluon density distributions in gold
nuclei for $0.001<x<0.1$, thereby extending our current knowledge and
including an overlap region that tests the universality of the gluon
distribution. 
\item
Characterization of correlated pion cross sections as a function of
$p_T$ to search for the onset of gluon saturation effects
associated with macroscopic gluon fields.
\item
Measurements with transversely polarized protons that are expected to
resolve the origin of the large transverse spin asymmetries in
$p_{\uparrow}+p\rightarrow \pi^0+X$ reactions for forward $\pi^0$
production.
\end{itemize}

In d+Au collisions, the FMS will face the deuteron beam and will see
neutral pions produced by large-$x$ quarks in the deuteron interacting
with the low-$x$ gluons in the Au nucleus.  The key to this analysis
is the detection of a second particle in coincidence with a triggering
particle in the FMS.  The coincident signal might be a high-$p_T$
track or jet detected in the TPCs or it might be a $\gamma$ or
$\pi^0$ detected in the BEMC or EEMC.  For $x<0.01$, the
coincident particle will be a second $\gamma/\pi^0$ detected in the
FMS, whose large acceptance makes this coincidence measurement
possible.  The spatial dependence of the nuclear gluon density
\cite{FSL,EKKV} will be investigated by analyzing two-particle
correlations as a function of particle multiplicity in the Au beam
direction with the existing STAR subsystems.

Analysis of the kinematics of the relative momentum between the
trigger particle and the coincident particle allows us to determine
$g(x)$ in the gold nucleus under the simple assumption of elastic
scattering of collinear initial-state partons.  This measurement of
the gluon density provides the essential input to the simulation codes
that attempt to determine the energy density achieved when heavy
nuclei collide, in the state which could expand to become the quark
gluon plasma.

The same correlated particle analysis will allow us to study the
physics of the parton saturation region, if it exists for $Q^2<4$
GeV$^2$.  This physics is associated with the transformation from a
parton-dominated picture of the nuclear gluon distribution to a
picture for which macroscopic QCD fields might play a role or provide
the most approporiate physics description.  The FMS granularity will
enable measurement of the azimuthal angle ($\phi$) of the trigger pion
and a coincident pion.  The peak in the
$\Delta\phi=\phi_{\pi1}-\phi_{\pi2}$ distribution at $180^{\circ}$,
the classic signature of parton elastic scattering, is expected to
broaden \cite{qiuvitev} or disappear \cite{KLM} ({\it ie.}, the
forward jet becomes a monojet and the recoil hadron rapidity
distribution is modified) when scattering from macroscopic gluon
fields dominates scattering from single gluons.

The analysis of FMS-triggered events at STAR will also be used in
polarized proton running where the extended calorimetry acceptance
will greatly enhance our ability to determine how quark and gluon
fields conspire to share the proton's $\frac{1}{2}$ unit of spin.
Polarized deep inelastic scattering (DIS) experiments found that the
intrinsic spin of quarks and antiquarks contribute only $\sim20\%$ to
the nucleon spin, contrary to early theoretical expectations of
$>60\%$.  A prime objective of the RHIC spin program is to understand
how gluon spin and parton orbital angular momentum play a role in this
``missing spin puzzle''.  The correlated pion analysis of the FMS and
the analysis channels it opens will play a crucial role in answering
these questions and could lead to resolution of the longstanding
question about the origin of the large transverse single spin
asymmetry in forward pion production.

\section{Nuclear gluon densities}
\label{gluon}

A central objective in high energy physics has been the
systematic characterization of parton (quark and gluon) density
distributions \cite{CTEQ,MRST}.  As a consequence of factorization
theorems we know that there is a class of high-$p_T$
two-parton (leading twist) experiments that can be understood in terms
of an initial state of independent partons within a proton. Within
this framework, the part of the cross section due to a particular
sub-process is equal to the product of a calculable parton-level cross
section and the two universal initial state parton probability
densities

$$\sigma(x_n,x_m) \propto \sigma_{nm} f_n(x_n) f_m(x_m).$$

The parton densities $f(x)$ are universal properties of the proton,
applicable in all hard scattering processes, and in most cases (but
not all) refer to the positive definite probability density to find a
parton ``$n$'' ($n=q$ for quark, $\overline{q}$ for antiquark and $g$
for gluon, with $f_g(x)$ referred to as $g(x)$) carrying a fraction
``$x$'' of the parent nucleon longitudinal momentum; $x$ is a
kinematic variable in DIS.  We combine contributions from all partonic
sub-processes that lead to the same final state and account for
contributions that come from all possible $x$ values by adding the
partial cross sections.

The nucleon gluon distribution $xg(x)$ is known in the region
$0.001<x<0.01$ (Fig.~\ref{fig:1}) but the nuclear distributions
are not.  Our present understanding of how parton distribution
functions (PDFs) are changed when nucleons are bound in a nucleus is
primarily derived from DIS of charged leptons from nuclear
targets. The charged leptons used in DIS interact with the
electrically charged quarks, not with the electrically neutral gluons,
and provide measurements of structure functions, $F_i$.  In the parton
model, $F_2(x,Q^2)=x\sum_{n} e_n^2[q_n(x,Q^2)+\overline{q}_n(x,Q^2)]$,
where $e_n^2$ is the squared electric charge of the quark of type $n$
and $Q^2$ is the squared four momentum transfer of the scattered
lepton, equated to the square of the scale at which the parton
substructures are observed.  Gluon densities are determined from the
QCD evolution equations \cite{DGLAP} applied to scaling violations of
$F_2$ measured for the nucleon over an
extremely large $x$ and $Q^2$ range at the HERA collider
\cite{H1_gluon,ZEUS_gluon}.  Sensitivity to $g(x)$ in DIS is
approximately given by the $Q^2$ variation of $F_2$ at half that $x$
value, $g(2x)\propto \partial F_2(x,Q^2) / \partial (\ln Q^2)$ \cite{Prytz}. The kinematic range
of the world data for the gluon distribution in nuclear targets is
shown in the right panel of Fig.~\ref{fig:1} as used in a recent global analysis
of nuclear modifications to PDFs
\cite{HKM,CERN_pA}. Such input to the nuclear gluon density is
available only for $x>0.02$ because nuclear DIS has been restricted to
fixed target experiments.  As will be discussed below, the study of
d(p)+Au collisions at $\sqrt{s_{NN}}$=200 GeV can provide direct
sensitivity to the nuclear modification of the gluon density for $x$
values on the order of $x \sim 0.001$ and can test the universality of
the nuclear gluon density in the range $0.02<x<0.1$.  Comparable
sensitivity in DIS to such low $x$ would require an electron-ion
collider \cite{eRHIC}. Measuring the PDF with quark and gluon probes
allows us to get to $x$ and $Q^2$ values where saturation phenomena
might be present.

\section{Tests of parton saturation}
\label{saturation}

Factorization theorems allow us to add cross
sections rather than quantum amplitudes, with partons considered
quantum mechanically independent of each another. Within this picture,
we are tempted to imagine that the gluon distribution of a nucleus
might be obtained by adding the gluon distributions for each nucleon,
with some accounting for the relative motion of the nucleons in the
nucleus. While perhaps true for large $x$ processes, at small $x$ the
uncertainty principle tells us that the partons will all overlap in
the longitudinal direction, so the partons do not interact
independently.  The front surface partons will interfere or shadow the
back surface partons of the nucleus. For more than 20 years it has
been recognized \cite{GLR} that the quantum independence of partons
cannot extend to very small $x$ where the gluon density is very large.

\begin{figure}
\begin{center}
\resizebox{0.6\hsize}{!}{\includegraphics{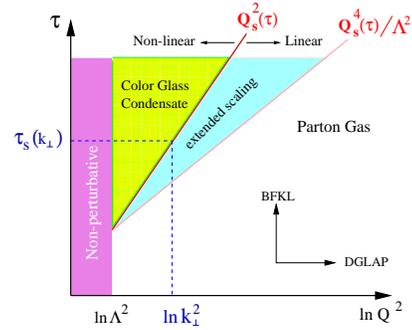}}
\end{center}
\caption{Diagram showing the boundary between possible ``phase''
  regions in the $\tau=\ln(1/x)$ versus $\ln(Q^2)$ plane
  \cite{IV}.  BFKL evolution \cite{BFKL} results in expected exponential
  growth of the gluon density at fixed $Q^2$ and increasing $\tau$.
  DGLAP evolution is discussed in the text.}
\label{phase}
\end{figure}

To determine the scale at which collective behavior might
become evident, the uncertainty principle suggests that a scattering
process at fixed $p_T$ will probe a transverse area approximately
given by $S(p_T)=\pi(\hbar / p_T)^2$ . For example at $p_T$ = 2 GeV/c,
this corresponds to about 0.3 mb, small in comparison to the proton
cross sectional area of about 30 mb. The number of gluons that are
present and that could shadow one another above $x$ is nominally given
by $n_{gluons}(x)=\int_x^1 g(x^{\prime})dx^{\prime}$.  At
$x$=0.01 $n_{gluons}\approx 7$ , increasing by 7--8 for each order of
magnitude decrease in the lower limit of $x$.  At $x$=0.001, the
product of cross section times number of gluons is $S(2$ GeV/c$)\times
n_{gluons}(0.001)\approx$ 5 mb.  This estimate suggests that for events
with these kinematics, the chances of finding more than one gluon
within the transverse resolution of the scattering probe is less than
20\%.  However in a nucleus of mass number $A$, the area of the
nucleus grows roughly as $A^{2/3}$ while the number of gluons would
nominally grow proportionally to $A$.  Thus, the transverse density of
nucleons should grow by a factor like $A^{1/3}$. For Au, this factor
of 6 in transverse density suggests that shadowing could be
substantial. By $x$=0.0001, it could become dominant. Of course, at
lower $p_T$, the effects would show up at a larger value of $x$.  Real
predictions for the onset of shadowing vary with the model used
\cite{FGS}, but whether shadowing modifies the gluon interactions at
RHIC is an experimental question which must be answered with data.

\begin{figure*}
\begin{center}
\resizebox{0.27\hsize}{!}{\includegraphics{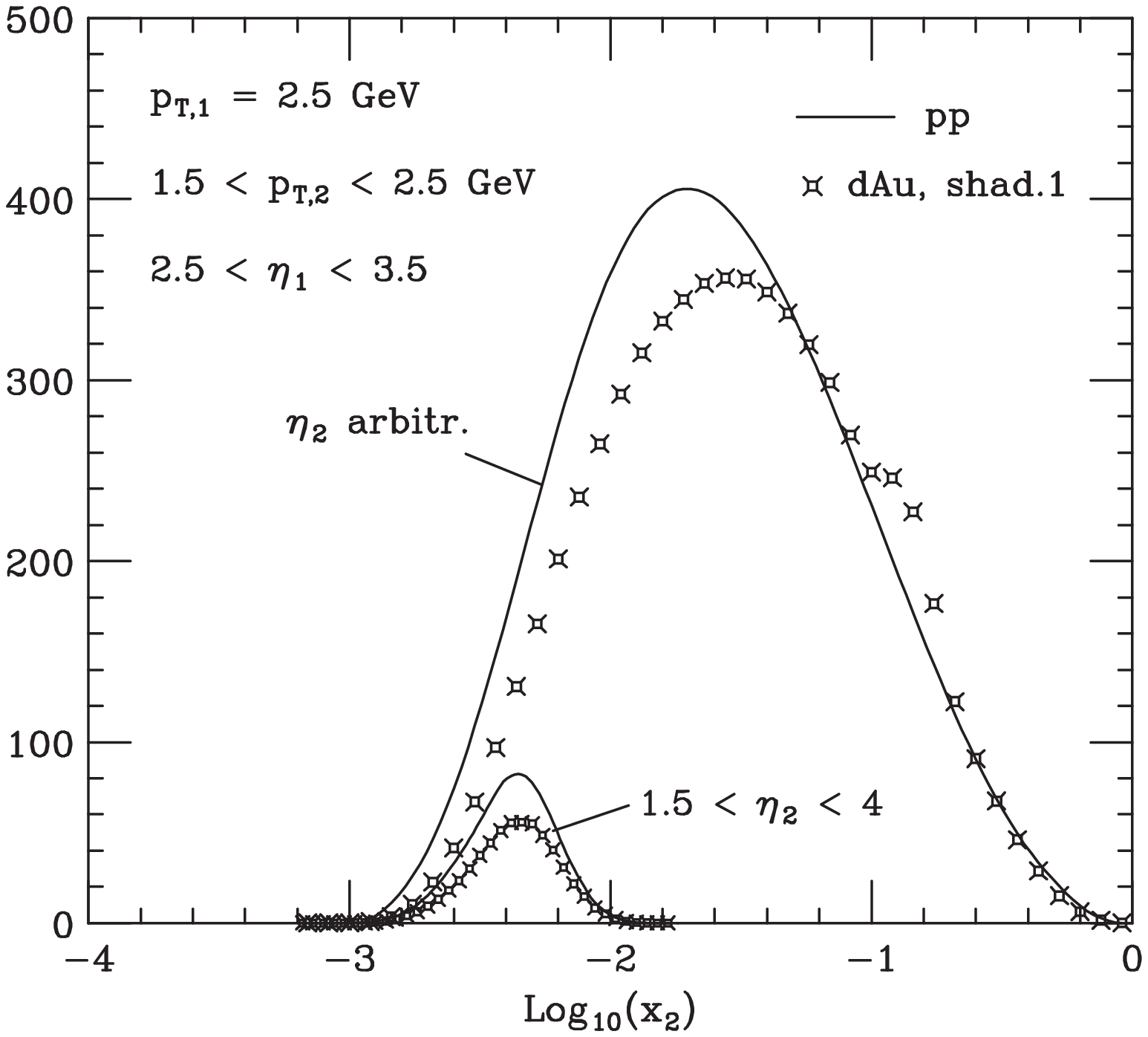}}
\resizebox{0.27\hsize}{!}{\includegraphics{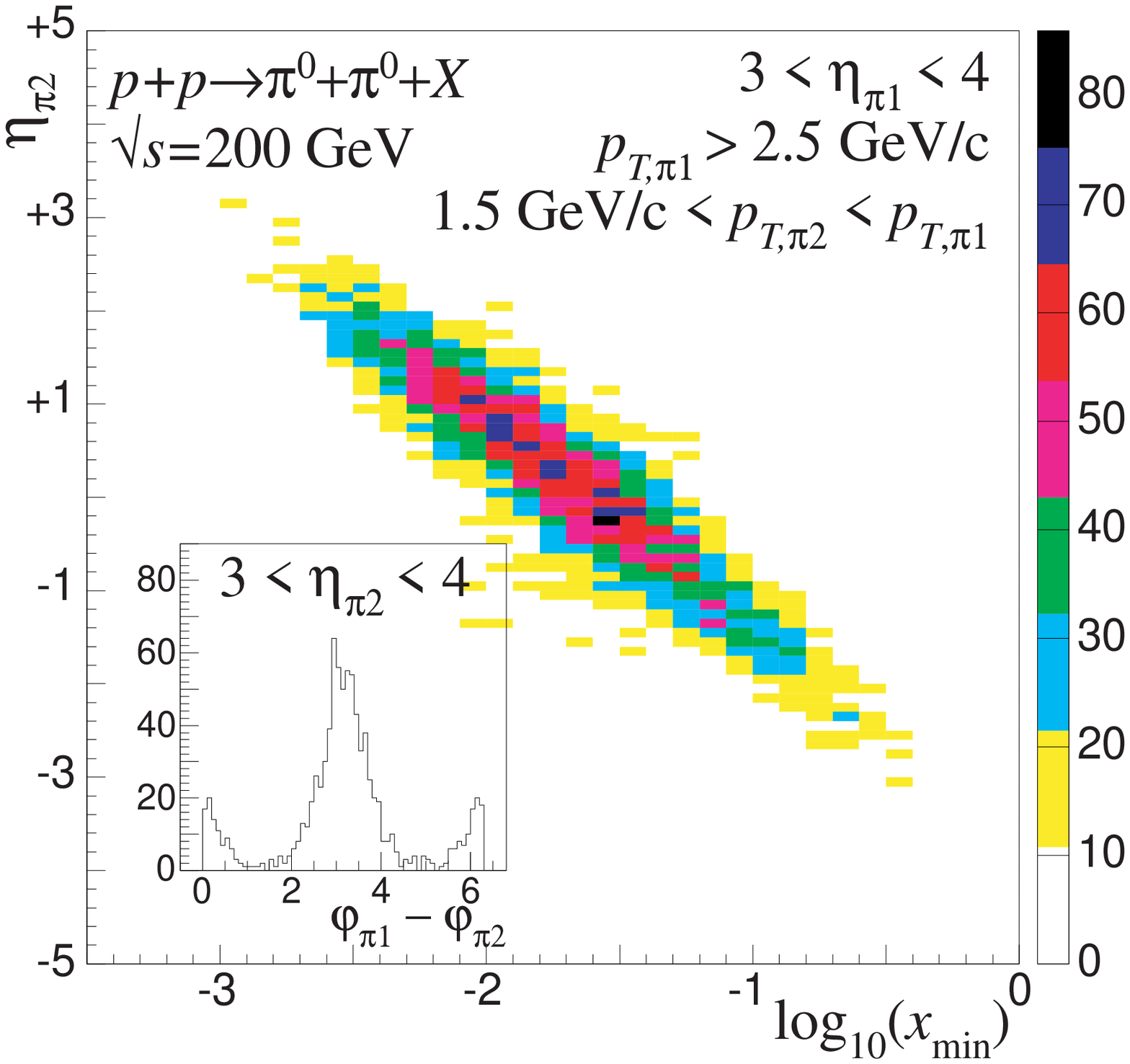}}
\resizebox{0.23\hsize}{!}{\includegraphics{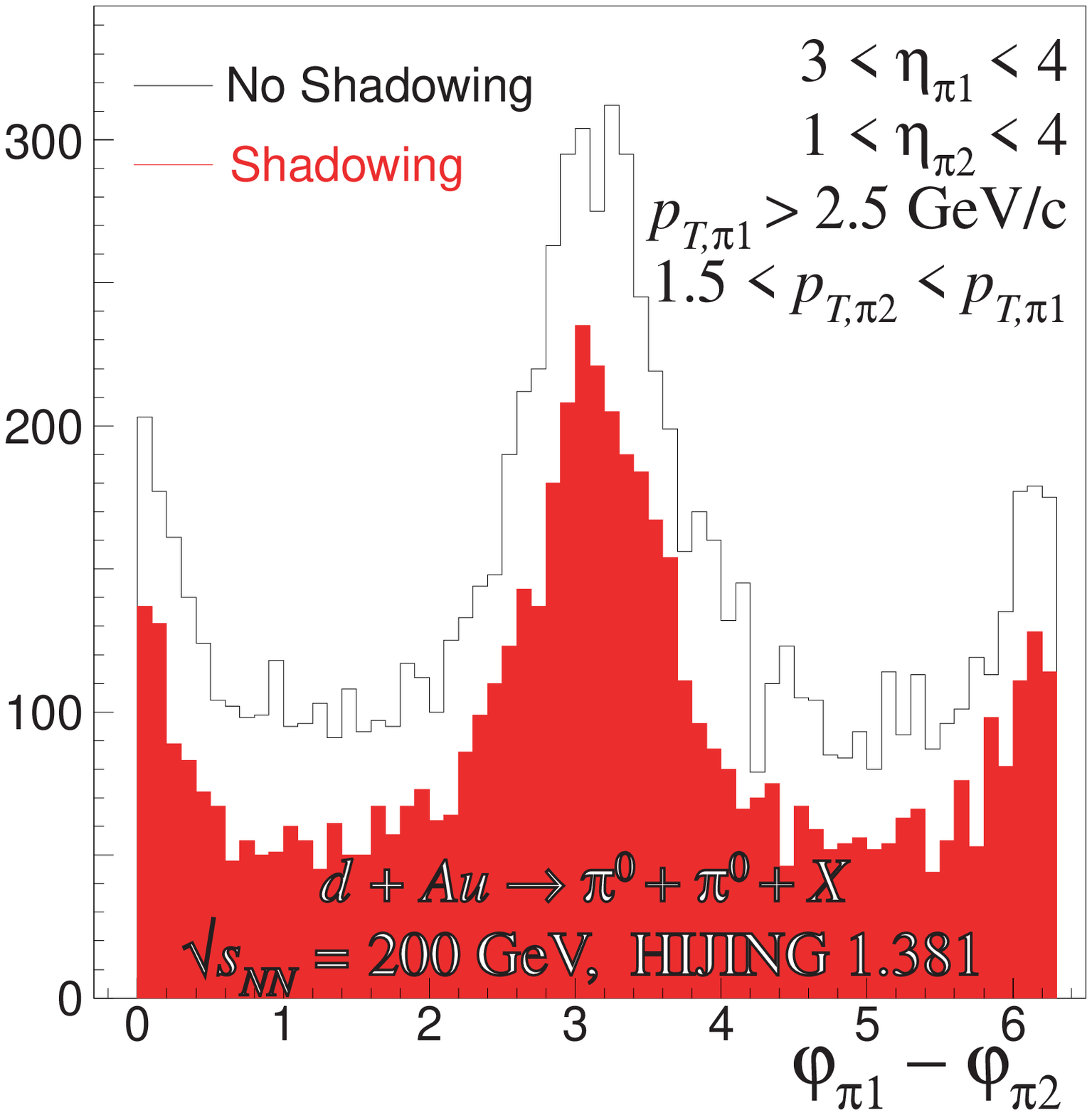}}
\end{center}
\caption{(Left) pQCD calculation of 
  $\pi^0-\pi^0$ production at large $\eta$ in p+p and
  d+Au collisions at $\sqrt{s_{NN}}$=200 GeV \cite{GSV}.  The
  distributions integrate to $d\sigma/dp_T$ in units of pb/GeV.  The
  smallest $x$ values are probed when $\pi^0-\pi^0$ pairs are detected
  at large $\eta$.  (Middle) PYTHIA \cite{pythia} simulation for
  $\pi^0-\pi^0$ production at large $\eta$ in p+p
  collisions at $\sqrt{s}$=200 GeV.  The $\eta$ of the associated
  $\pi^0$ is strongly correlated with the $x$ value of the soft parton
  involved in the partonic scattering. (Right)  HIJING simulation for
  $\pi^0-\pi^0$ production at large $\eta$ in d+Au
  collisions at $\sqrt{s}$=200 GeV.  Compared to the p+p
  simulations, the peaks in $\Delta\phi=\phi_{\pi1}-\phi_{\pi2}$ corresponding to
  elastic parton scattering, sit atop a background from other
  mechanisms for particle production.}
\label{fig:2}       
\end{figure*}

For p+p collisions at $\sqrt{s}$=200 GeV, unlike at lower $\sqrt{s}$ \cite{BS},
next-to-leading order (NLO) pQCD calculations \cite{AJF} quantitatively
describe inclusive particle production down to $p_T$ of $\sim$2 GeV/c using
PDFs \cite{CTEQ,MRST} and fragmentation functions
\cite{KKP,kretzer} that describe the hadronization of the scattered partons.
Furthermore, di-hadron
azimuthal correlations have the same structure at these moderate $p_T$ 
as they do at the highest possible $p_T$ values, consistent with
the idea that elastic scattering of quarks and gluons is responsible
for the particle production \cite{ogawa}.

Recent measurements at STAR using the prototype
FPD already indicate that the factorized leading twist 
pQCD calculations work quite well to predict the $p+p\rightarrow
\pi^0+X$ cross section in the $3<\eta< 4$ region
\cite{STAR_FPD}. This gives
confidence in the interpretation that at $\sqrt{s}$ =200 GeV the
particle production process is dominated by leading twist quark- gluon
scattering. With the FMS focusing on $\pi^0-\pi^0$ pairs, we will
select the low-$x$ component shown by pQCD calculations 
(Fig.~\ref{fig:2}) to make only small contributions to the inclusive
measurement.  The low-$x$ component of the forward pion yield is where
shadowing effects are expected to be most important \cite{vogt}.  In
the middle panel of Fig.~\ref{fig:2} we see that when triggering on a
$\pi^0$ in the range $3<\eta<4$, the rapidity of the second $\pi^0$
will reflect the $x$ of the struck gluon. The right panel of
Fig.~\ref{fig:2} shows that elastic parton scattering is identified
above physics backgrounds in d+Au collisions.

There has been considerable recent interest among the experts in the
application of pQCD in reconciling the meaning of shadowing with the
idea of universal (factorizable) PDFs. What has
emerged recently \cite{BHMPS} is a better understanding of just what the
universal parton density means at small $x$.  The present understanding
is that the ``shadowed'' small $x$ distributions should be universal but
do not strictly reflect the probability for finding a parton. Included
in the universal factorized functions are built-in final-state
correlations with other gluons in the proton.  We now see that even
from the strict, leading twist perspective, low-$x$ perturbation theory
has a different interpretation from large $x$ because it always involves
a sampling of the macroscopic gluon fields. In light of this, there is
real excitement that a variety of low-$x$ phenomena from shadowing to
large transverse spin asymmetries may be tied together with pQCD in
ways not before appreciated.

Measurements of gluon shadowing at RHIC and the LHC will be essential
input for models that predict the relationships between quark
distributions and macroscopic gluon fields. Among the descriptions of
shadowing or saturation effects is the Color Glass Condensate (CGC)
\cite{GLR,MQBM,MV}, an effective-field theory for understanding parton
saturation.  In the CGC picture, the saturation effects are associated
with a new phase of the gluon field. The onset of this phase can
be probed by measurements at small $x$ and at small $Q$ (related to
the produced parton mass and the $p_T$ associated
with the scattering).  Mapping out the boundaries (Fig.~\ref{phase}) for saturation
signatures for back-to-back jet correlations as a function of $x$ and
$p_T$ is a primary mission of the FMS.

\section{Early results from RHIC}
\label{results}

The first Au+Au collision runs at $\sqrt{s_{NN}}$=200 GeV resulted in
the observation that high-$p_T$ particle production at midrapidity was
suppressed relative to expectations resulting from the scaling of
yields from p+p collisions. The data also showed that two-particle
correlations were suppressed when the particles were back-to-back
(away-side, $\Delta\phi\approx180^{\circ}$) but not when they were
fragments of the same jet (near-side,$\Delta\phi\approx0$ or 2$\pi$) \cite{STAR_btob}.  These
observations are consistent with a prediction based on radiative
energy loss of a high-$p_T$ parton passing through a quark-gluon plasma
\cite{GPWG}. A d+Au run at $\sqrt{s_{NN}}$ =200 GeV was scheduled early in the RHIC
program to eliminate the possibility that this was due to
initial state effects.  For d+Au collisions, midrapidity particle
production was found to have a small enhancement, consistent with the
Cronin effect \cite{Cronin}, and back-to-back correlations \cite{STAR_dau} more closely
resembled results from p+p collisions than from Au+Au collisions.
Hence, the suppression of back-to-back correlations in Au+Au
collisions was attributed to the strongly interacting matter formed in
those collisions, matter that was formed by interactions of
the low-$x$ gluons.  

\begin{figure*}
\begin{center}
\resizebox{0.25\hsize}{!}{\includegraphics{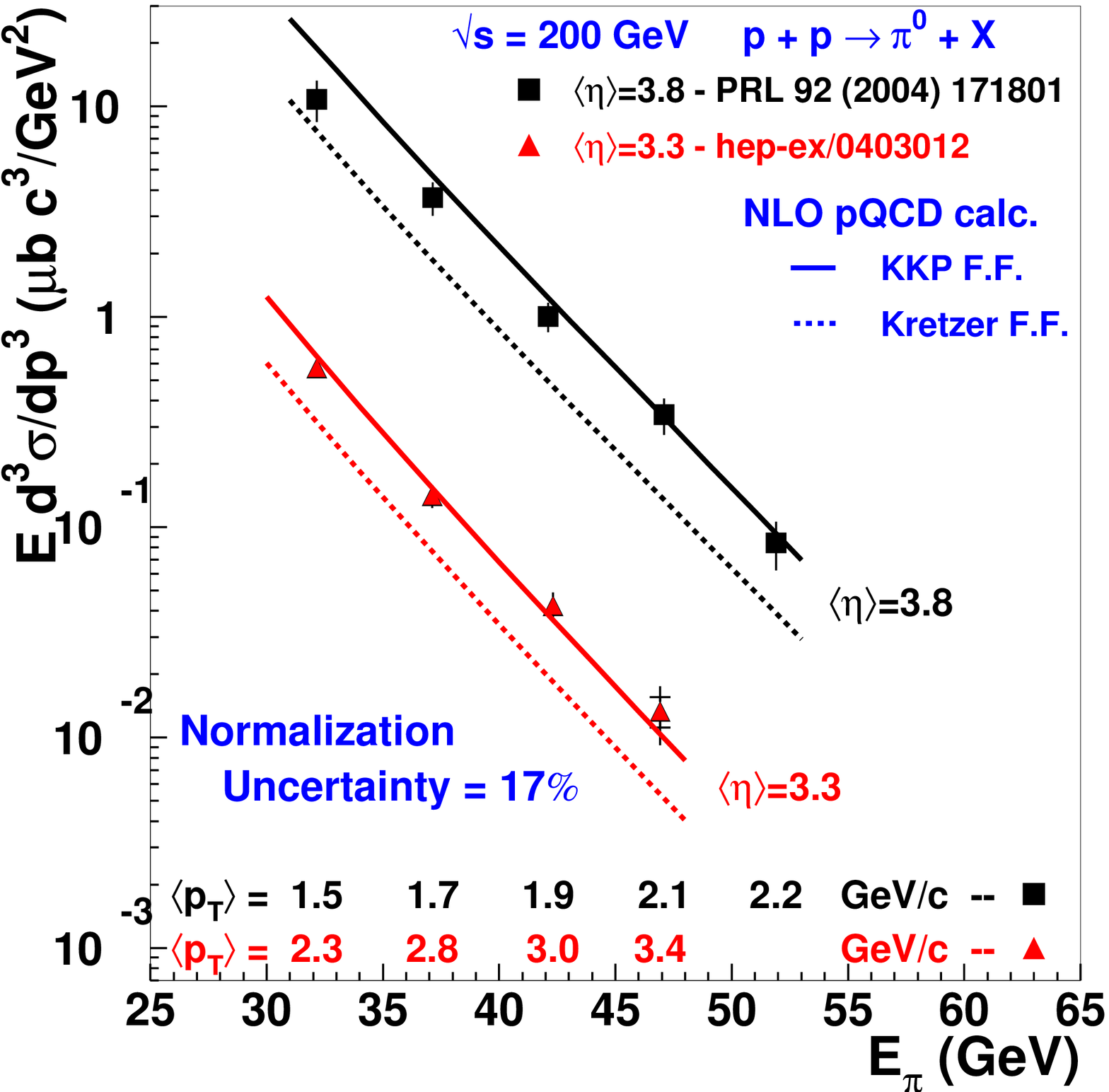}}
\resizebox{0.25\hsize}{!}{\includegraphics{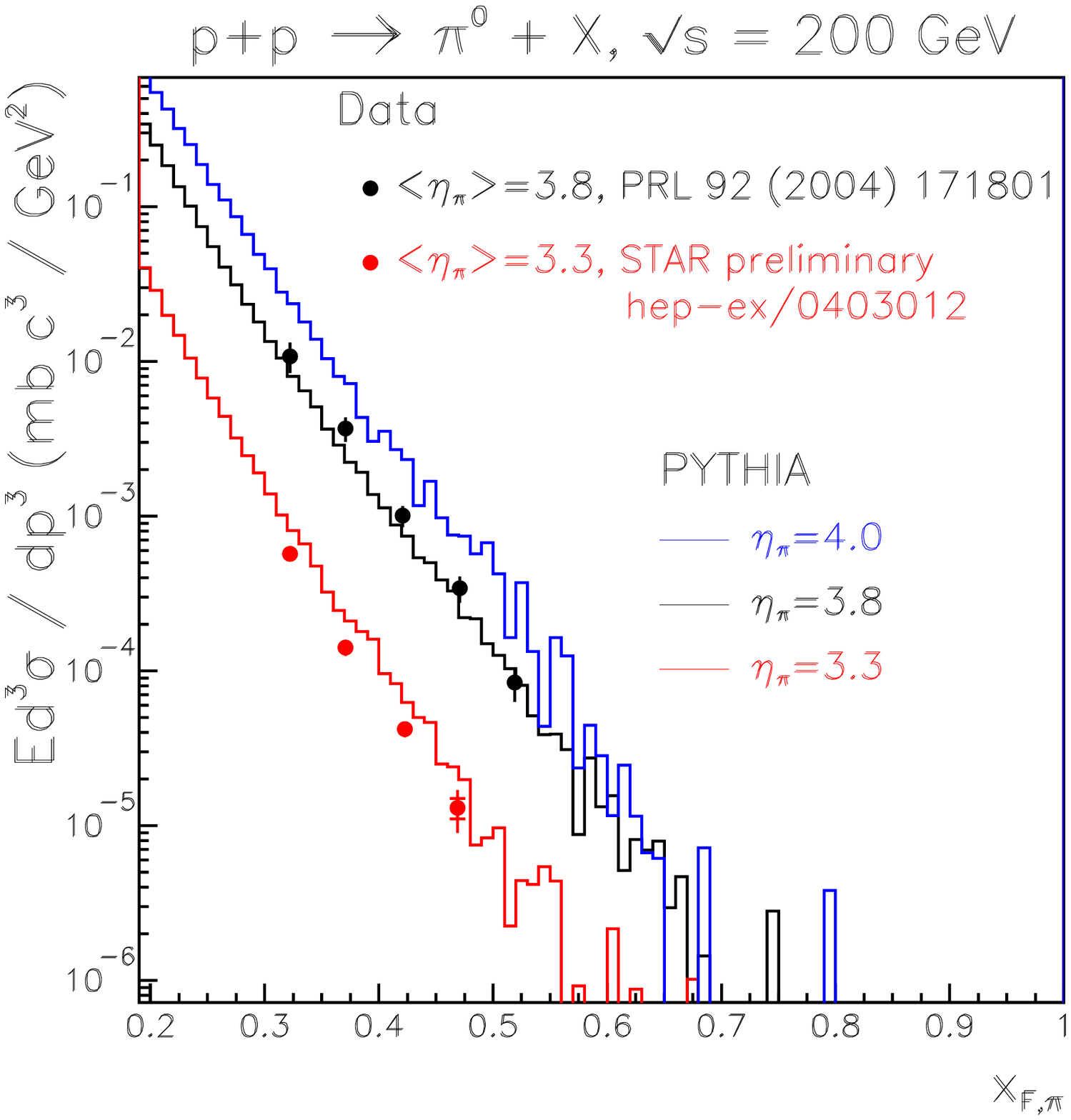}}
\resizebox{0.25\hsize}{!}{\includegraphics{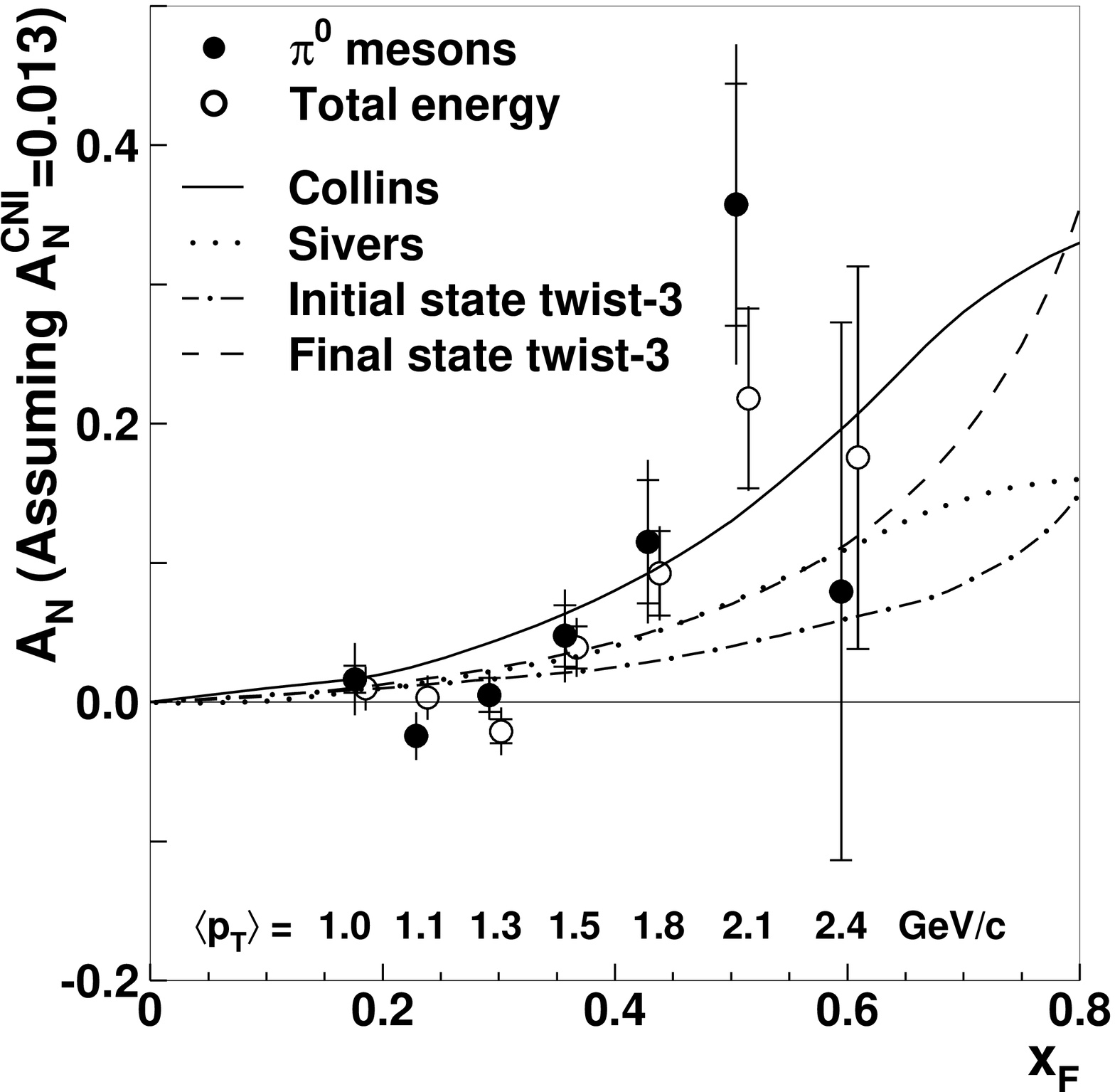}}
\end{center}
\caption{(Left) Invariant cross sections for inclusive $\pi^0$
  production at large rapidity in p+p collisions at $\sqrt{s}$=200 GeV
  \cite{STAR_FPD,dubna} compared to NLO pQCD calculations
  \cite{AJF,KKP,kretzer}.  (Middle) The data are compared with
  predictions from PYTHIA \cite{pythia}.  (Right)  The analyzing power
  for $\pi^0$ production at $<\eta>$=3.8 in p+p collisions at
  $\sqrt{s}$=200 GeV \cite{STAR_FPD}.  The curves are predictions from
  pQCD models \cite{twist3,sivers_anselmino,collins_anselmino} evaluated at $p_T$=1.5 GeV/c.}
\label{fig:3}       
\end{figure*}

Measurements made by the BRAHMS collaboration
for d+Au collisions \cite{BRAHMS} showed that inclusive particle production
was suppressed as the rapidity of the observed
particles increased.  This provided a hint that the gluon distribution
in the Au nucleus may be depleted at low-$x$.  It is easy to understand
how this suppression can occur within the standard pQCD picture of
particle production.  In that picture, the quarks and gluons each
carry a fraction of their parent hadron momentum given by $x$. They
elastically scatter and then fragment to the final state hadrons
observed with a given $p_T$ and at a given $\eta$.
For collinear parton pairs, it is easily shown that 
$$x_+\approx \frac{p_T}{\sqrt{s}}(e^{+\eta_1}+e^{+\eta_2})\rightarrow x_F$$
$$x_-\approx \frac{p_T}{\sqrt{s}}(e^{-\eta_1}+e^{-\eta_2})\rightarrow 
x_F e^{-(\eta_1+\eta_2)},$$
where the Feynman-$x$ variable is $x_F=2p_L/\sqrt{s}$, relevant in
the limit $\eta_1 >> \eta_2$, and $p_L$ is the
longitudinal momentum component of the large $\eta$ particle.  For
inclusive particle production, one of the two jets, or its hadronic
surrogates, is observed at
$\eta_1$ and the second jet has a broad
$\eta_2$ distribution and $\Delta\phi\approx180^{\circ}$.  By
detecting a high energy hadron at large $\eta_1$, initial states with a
large-$x$ parton (most probably a quark) and a low-$x$ parton (most
probably a gluon) are selected.  For each unit rapidity increase, the
average $x$ of the gluon from the initial state parent hadron is
decreased by $e$.  A similar decrease in $x$ is a consequence of
studying particle production in collisions at higher $\sqrt{s}$.  
Hence, the observed suppression of particle
production at increasing rapidity can be interpreted as a reduction in
the probability of finding gluons in the nucleus at small $x$.

The BRAHMS results \cite{BRAHMS} were confirmed in measurements by
PHENIX \cite{PHENIX}.  The STAR collaboration also made measurements
of large rapidity particle production and produced a limited data
sample for two-particle correlations involving a large rapidity
$\pi^0$ \cite{ogawa}, measured with the FPD.  The
topology of the energy deposition in the FPD allows for robust
measurements of the energy and direction of neutral pions.  Since the
$\pi^0$ is a pseudoscalar particle, kinematic distributions of its
diphoton decay are exactly calculable in any frame of reference.
Hence, calibrations of the FPD response can be obtained at the level
of $\sim$1\% simply by requiring a fully consistent response of all
cells of the calorimeter to the photons produced by the
$\pi^0\rightarrow\gamma\gamma$ decay.  This same technique will be
used for the FMS.  Figure \ref{fig:3} also shows that the simulation code
PYTHIA \cite{pythia} is able to reproduce the absolute cross section
of the produced $\pi^0$ from p+p collisions.  In addition, nearly all
features of the $\Delta\phi$ distributions for charged hadrons with
$|\eta|<0.75$ coincident with forward $\pi^0$ are reproduced by PYTHIA.  This
gives us a tool to guide the design of the FMS and the interpretation
of the data.

\section{Proton spin with the FMS}

Our understanding of the two-particle correlations involving large
rapidity particles, and our ability to use these correlations to
measure nuclear gluon distributions, is a direct result of the
methodology developed to understand the first spin asymmetry
measurements at RHIC. It is no surprise that the FMS will also be a
powerful tool in studying the spin structure of the proton.

An early prediction of pQCD was that, at leading twist and with
collinear factorization, the chiral properties of the theory would
make the analyzing power($A_N$) small for particles produced with
transversely polarized proton beams \cite{kane}. $A_N$ is derived from
the azimuthal asymmetry of particle yields from a transversely
polarized beam incident on an unpolarized target.  However, from AGS
energies \cite{lowerenergy} to Fermi Lab energies \cite{e704} and most
recently at STAR \cite{STAR_FPD} (Fig.~\ref{fig:3}), a large transverse single spin
asymmetry has been observed.  The consistent trend is that
$A_N$ in $p_{\uparrow}+p\rightarrow \pi^0+X$ increases rapidly for
$x_F$ above about 0.3.  Transverse single spin asymmetries have
also been observed in semi-inclusive DIS from polarized targets
\cite{HERMES} and experimental studies of these spin effects is an
active area of research.  The FMS is ideally suited to extend these
studies.  Calculations of twist-3 contributions \cite{twist3} to the
observed $A_N$ provide terms that may be related to macroscopic
gluon fields in the polarized nucleon.  

There are multiple
phenomenological effects that have been identified as possible sources
for the large $A_N$, but only two that are expected to be
large. One is the Sivers effect \cite{sivers,sivers_anselmino}, which
is an initial state correlation between the parton intrinsic
transverse momentum $k_T$ and the transverse spin of the nucleon,
$A_N\propto {\bm S}_T \cdot ({\bm P}\times{\bm k}_T)$, where ${\bm P}$
is the beam momentum and ${\bm S}_T$ is the transverse proton spin. In
the Sivers framework, $A_N$ is sensitive to the contribution of quark
orbital angular momentum to the nucleon spin. Large $A_N$
is the result of a spin dependent $p_T$ trigger bias
favoring events where $k_T$ is in the same direction as $p_T$.

If the Sivers effect is present, we can further characterize it
with a measurement of the away side jet. The $k_T$ of the
initial-state parton can be measured when final-state jet pairs are
not exactly back to back ($\Delta\phi=180^{\circ}$)
\cite{boer_vogelsang}. The spin dependence of this $k_T$ measurement is
exactly what the Sivers model predicts.

While the Sivers effect connects $A_N$ to the orbital angular momentum of
quarks, a second effect, called the Collins effect
\cite{collins,collins_anselmino}, is directly sensitive to the
transversity distribution function \cite{RS,JJ}, related to the transverse
polarization of quarks (and antiquarks) in a transversly polarized
proton \cite{BDR}.  Here the quark scatters, preserving its transverse spin, and
then fragments into pions and other hadrons. The fragmentation
function reveals the polarization of the fragmenting quark and thus
the initial quark state.  In this example, the asymmetry does not
appear in the jet production directly, but only in the fragmentation.
The jet axis would not show the transverse asymmetry, but a pion
fragment would.  Recent calculations that include the full $k_T$
dependence in the convolution integrals provide some indication
that the Collins effect may be small \cite{ABDFM}.

The FMS will be able to distinguish between these mechanisms. By
looking at pairs of same-side neutral pions, we can measure
$A_N$ as a function of the two pion kinematics. With the FMS we
will separately measure the contributions to $A_N$ that comes
from the jet axis versus that which comes from the jet structure. Many
theory papers have studied this problem: however, the need for data is
great.  The FMS STAR experiments on transverse polarization will
provide to theorists the necessary input to determine the relative
contributions from the Sivers effect and the Collins effect.

\begin{figure}
\begin{center}
\resizebox{0.65\hsize}{!}{\includegraphics{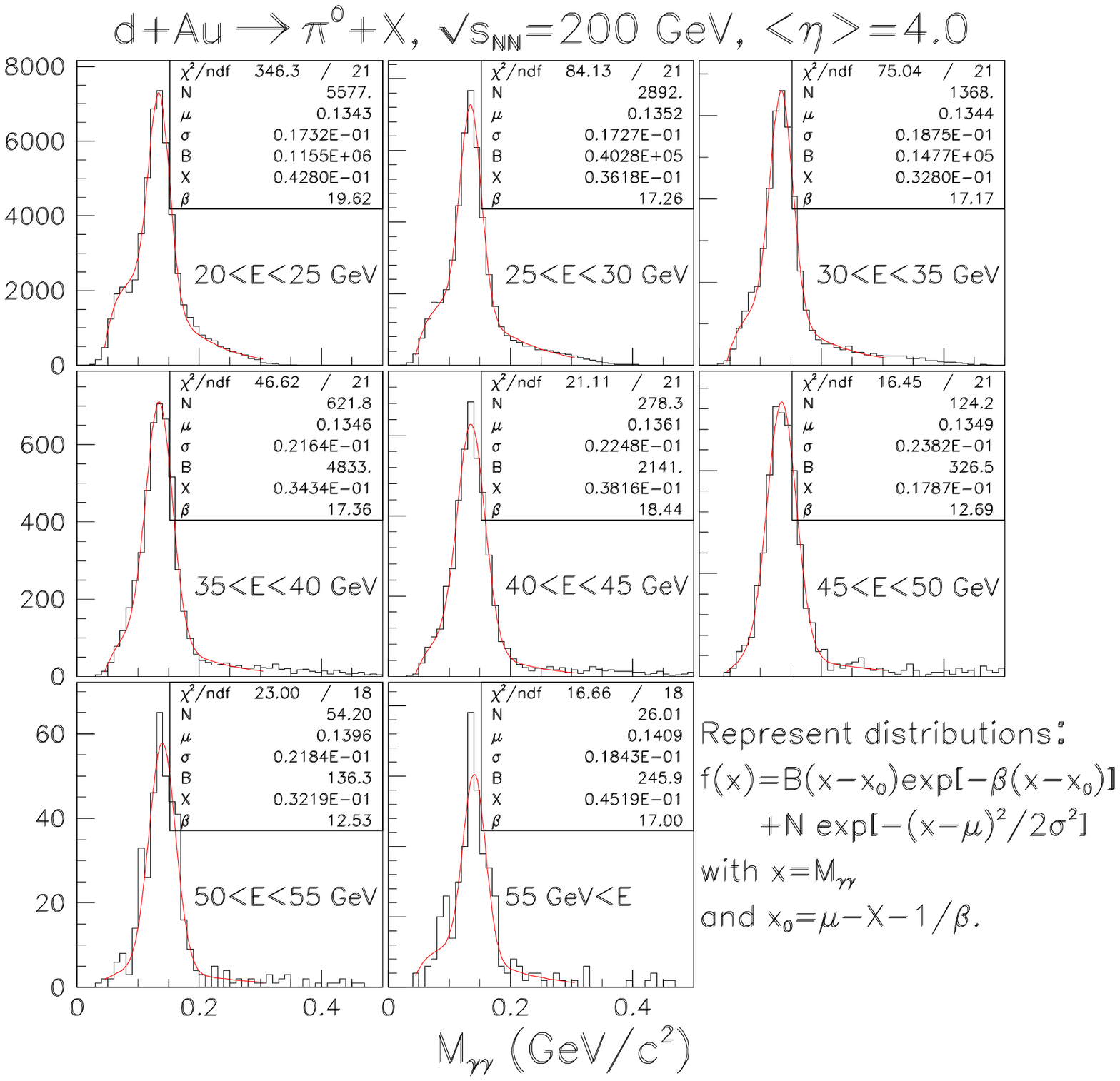}}
\resizebox{0.55\hsize}{!}{\includegraphics{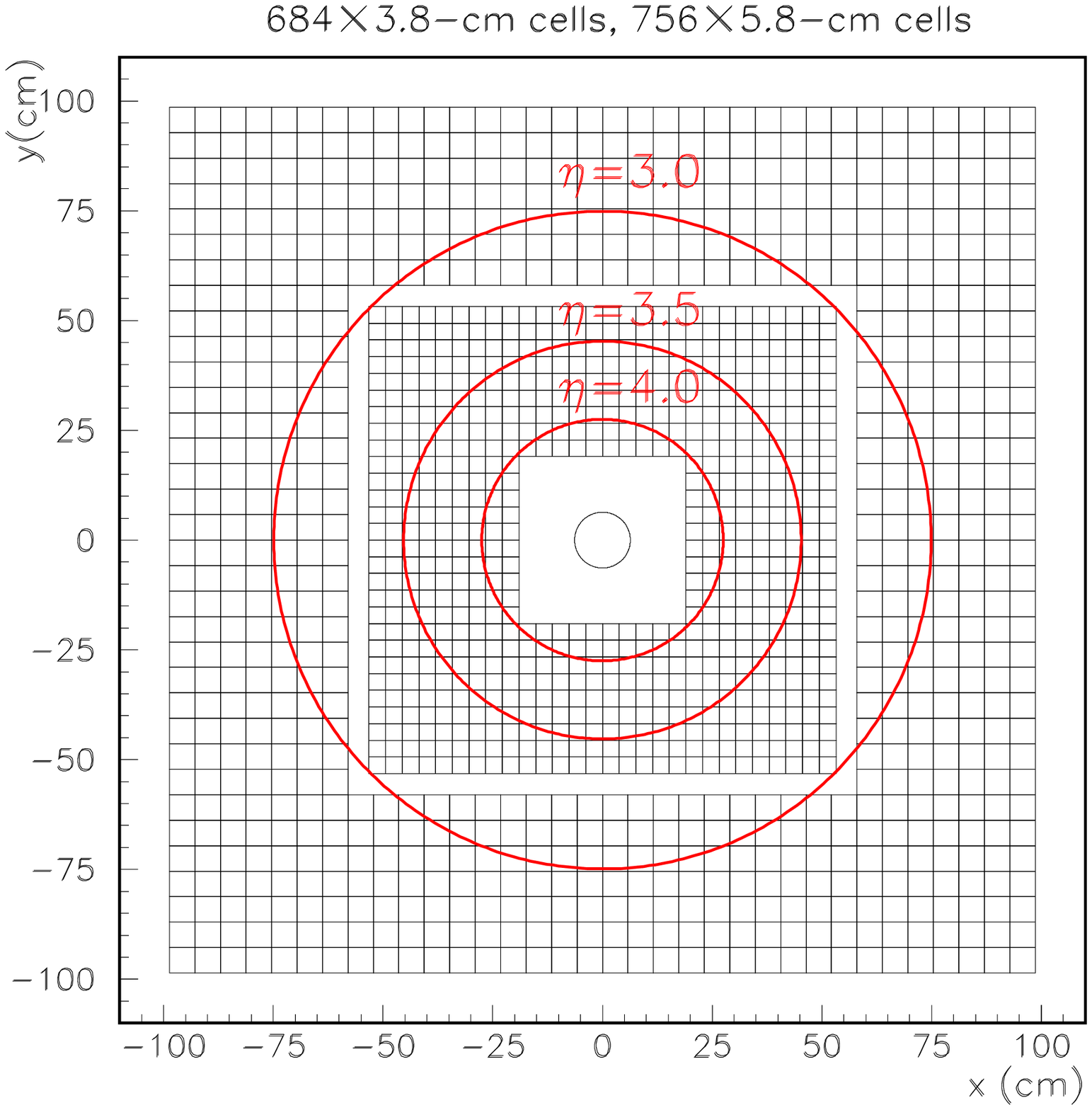}}
\end{center}
\caption{(Top) Di-photon invariant mass distributions from the STAR
  FPD for d+Au collisions at $\sqrt{s_{NN}}$=200 GeV at $<\eta>$=4.0 relative to
  the incident deuteron beam.  (Bottom) Layout of the proposed FMS.}
\label{config}
\end{figure}

\section{FMS Configuration}

The STAR FPD now taking data acts as the
prototype for the proposed FMS.  Our results
with the FPD demonstrate the feasibility of large rapidity
measurements with electromagnetic calorimetry in both p+p 
and d+Au collisions at $\sqrt{s_{NN}}$=200 GeV at RHIC. Each FPD
calorimeter is a $7\times7$ matrix of 3.8cm $\times$ 3.8cm $\times$
45cm lead-glass cells that can be positioned in
the range $3.3 < \eta < 4.0$. These are identical to the 684 small
cells to be used in the FMS and the techniques we have developed for
FPD tuning and analysis are directly applicable to the FMS.  By
implementing the FMS in STAR, the study of $\gamma-\pi^0$ and
$\pi^0-\pi^0$ correlations in both rapidity and azimuthal angles is
enabled over $-1<\eta<4$.  The FMS also
allows inclusive measurements of $\pi^0$, $\gamma$ and the production
of heavy mesons that decay to all $\gamma$ final states over
a broad range of $\eta$ and $p_T$. A schematic of the proposed FMS
detector is shown in Fig.~\ref{config}, along with examples of the mass
resolution from a topological analysis \cite{lednev} of the energy
deposition for data obtained with the current FPD. We expect to
achieve 1\% accuracy in the calibration of the FMS, just as we have
done for the FPD. Energy resolution of $<15\%/\sqrt{E}$ has been
demonstrated.  Simulation studies compare well to FPD data and show
that we can expect to locate the $\pi^0$ to better than 0.5 cm
(RMS). Our expected $\pi^0$ mass resolution is $\sim$23 MeV/c$^2$
based on experience with the FPD. For neutral pions with $20<E<60$ GeV
the reconstruction efficiency of the FPD is just given by geometry.
To predict event yields, we used a very conservative estimate of 35\%
for reconstruction and geometric efficiency.  The FMS will provide
complete azimuthal coverage for the pseudorapidity interval
$2.5<\eta<4.0$ and will be built from existing lead-glass cells.  

\section{Plan for measurements}

We will concentrate on measurements directed at our three immediate
physics goals for d+Au and for p+p running. In a future d+Au run at
$\sqrt{s}$=200 GeV we will determine the gluon distribution in the
gold nucleus. The large acceptance of the FMS will provide good statistics for
comparisons of identified $\pi^0$ yields in p+p and d+Au collisions over a
very broad range in $p_T$ and pseudorapidity.
We will measure the correlations between a trigger
particle in the FMS (either a $\pi^0$ or a $\gamma$) and a second
particle (jet surrogate) in the TPC, BEMC, EEMC or FMS.
The threshold $p_T$ for the
trigger and coincident particles will be investigated
over the range from $1<p_T<4$ GeV/c. The large size and high
granularity of the FMS allow us to identify neutral pions at energies
down to $\sim$10 GeV, where hadronic energy deposition becomes a
significant background.  At maximum rapidity, this determines our
lowest observable $p_T$ threshold.  As seen in Fig.~\ref{fig:2}, the $\eta$ of
the coincident particle correlates with the gluon $x$ value. The
$\Delta\phi$ distribution reveals the
scattering history. To develop the full picture of the gluon
distributions, we will investigate the dependence of $\pi^0-\pi^0$
correlations on $Q^2$.  To determine the dynamical origin (Sivers or Collins) of the
observed transverse single spin asymmetry for forward $\pi^0$
production, we will make similar correlation measurements in the
polarized p+p run.

As shown in Fig.~\ref{fig:2}, pQCD calculations \cite{GSV} suggest that the
lowest $x$ values for the gluon density are probed when both jets from
the elastic parton scattering are produced at large rapidity.  Both
$\pi^0+\pi^0$ and $\pi^0+\gamma$ final states can be analyzed in
the FMS. The $p_T$ of the $\pi^0$ must be large enough to favor
elastic parton scattering over inelastic scattering \cite{ogawa}
although both contributions are contained in PYTHIA and NLO pQCD
calculations. The large size of the proposed FMS enables the use of
isolation cuts to distinguish between $\pi^0$ decay photons and direct
photons. The dominant subprocess for direct photon production at RHIC
is QCD Compton scattering, $qg\rightarrow \gamma q$. This can also be
used to probe the small $x$ gluon density with only minimal physics
backgrounds.

Figure \ref{fig:2} shows a HIJING 1.381 \cite{hijing}
simulation of $7.5\times 10^8$ minimum-bias d+Au events at
$\sqrt{s_{NN}}$=200 GeV, from which all $\pi^0-\pi^0$ pairs with the
specified $p_T$ and $\eta$ are selected and used to compute
$\Delta\phi$.  Unlike the case for p+p collisions, the elastic parton
scattering peaks in the $\Delta\phi$ distribution sit atop a
background from the nuclear collision.  Despite the background, the
elastic parton scattering is readily discriminated and can be
identified as the expected peak in $\Delta\phi$.  We can
quantitatively describe these distributions from $1< \Delta \phi
<5.28$ radians by a Gaussian function, used to model the peak, and a
constant background.  A best fit to the $\Delta\phi$ distribution in
Fig.~\ref{fig:2} results in $2.3\times 10^3$ events in the peak for simulations
done without shadowing and $\sim 1.8\times 10^3$ events in the peak
for the simulations done with shadowing.  We can expect that a 10-week
d+Au run will allow us to sample $>6\times10^{10}$ minimum-bias
interactions, based on RHIC performance for d+Au collisions achieved
in the last weeks of run 3.  Accounting for detector efficiencies for
the FMS and endcap EMC, the simulations suggest we will observe at
least $8\times 10^3$ events in the $\Delta\phi$ peak without
shadowing.  This is enough to investigate the spatial dependence of
the nuclear gluon density \cite{FSL,EKKV} using particle multiplicity
measurements in the Au beam direction made by other STAR subsystems to
determine sensitivity to the impact parameter of the collision.
Broadening or disappearance of the away-side correlation could signal the
transition to a macroscopic gluon field at sufficiently low $x$.

For the polarized proton running, we base estimates of the rate of
near-side $\pi^0-\pi^0$ pairs on PYTHIA \cite{pythia}.  The STAR
measurements (Fig.~\ref{fig:3}) show that $A_N$ is small below $x_F \sim 0.4$,
and increases monotonically as $x_F$ of the forward $\pi^0$ increases.
In the Sivers picture, $A_N$ should be associated with the forward jet
and should be present for $\pi^0-\pi^0$ pairs from the same jet.  We
would expect a large $A_N$ when $x_{F1}+x_{F2} > 0.4$.  When one
$\pi^0$ is observed at $3 < \eta < 4$ with $x_{F1} > 0.25$ and a
second $\pi^0$ is observed with $\eta < 4$ and $x_{F2} > 0.15$ and the
$\pi^0-\pi^0$ pair has $|\eta_1-\eta_2| < 0.5$, the simulated
$\Delta\phi$ correlation shows a jet-like near-side correlation peak
sitting atop a uniform background attributed to the underlying event.
We expect $\sim1.5\times10^4$ $\pi^0-\pi^0$ events in the near-side
jet-like peak for 1 pb$^{-1}$ of integrated luminosity for polarized
p+p collisions at $\sqrt{s}$ = 200 GeV with the FMS.  For a beam
polarization of 50\% this would result in a statistical error on
the analyzing power of $\delta A_N \sim$ 0.01.  The Collins mechanism
attributes $A_N$ to the correlation between the momenta of two hadrons
from the same jet and the proton spin vector.  The transverse momentum
associated with jet fragmentation producing a $\pi^0$ with
$x_{F1}>0.4$ and $3 < \eta < 4$ can be measured by detecting a second
$\pi^0$ with $\eta < 4$ and $x_{F2}>0.15$ and requiring the
$\pi^0-\pi^0$ pair have $|\eta_1-\eta_2| < 0.5$.  Again, the
$\Delta\phi$ correlation shows a jet-like near-side correlation peak
sitting atop a uniform background.  For these kinematics, we expect
$2\times10^3$ $\pi^0-\pi^0$ pairs in the jet-like near-side
$\Delta\phi$ peak for 1pb$^{-1}$ of integrated luminosity.  If a
non-zero Collins effect is observed, then larger integrated luminosity
samples would be required to map out the $x$ dependence of the
transversity structure function.

\section{Summary}

Early experimental results from RHIC, coupled with theoretical
developments that demonstrate the robustness of NLO pQCD calculations at
collider energies down to moderate $p_T$ values,
suggest that a quantitative determination of the gluon density in a
heavy nucleus can be obtained at an order of magnitude lower $x$ than
available from DIS on nuclear targets.  Such studies require improved
instrumentation in the forward direction at RHIC.  In addition to
providing crucial information to understand the initial state of
heavy-ion collisions that may lead to a quark-gluon plasma, future low-$x$
studies at RHIC may establish the existence of a macroscopic gluon
field.  Improved forward instrumentation can also disentangle the
dynamical origin of transverse single spin asymmetries.

%
%
%
%
%

\end{document}